# Origin of the mosaicity in graphene grown on Cu(111)

Shu Nie[1], Joseph M. Wofford[2], Norman C. Bartelt[1], Oscar D. Dubon[2], and Kevin F. McCarty[1, ‡]

[1] Sandia National Laboratories, Livermore, CA 94550

[2] Department of Materials Science & Engineering, University of California at Berkeley

and Lawrence Berkeley National Laboratory, Berkeley, CA 94720

We use low-energy electron microscopy to investigate how graphene grows on Cu(111). Graphene islands first nucleate at substrate defects such as step bunches and impurities. A considerable fraction of these islands can be rotationally misaligned with the substrate, generating grain boundaries upon interisland impingement. New rotational boundaries are also generated as graphene grows across substrate step bunches. Thus, rougher substrates lead to higher degrees of mosaicity than do flatter substrates. Increasing the growth temperature improves crystallographic alignment. We demonstrate that graphene growth on Cu(111) is surface diffusion limited by comparing simulations of the time evolution of island shapes with experiments. Islands are dendritic with distinct lobes, but unlike the polycrystalline, four-lobed islands observed on (100)-textured Cu foils, each island can be a single crystal. Thus, epitaxial graphene on smooth, clean Cu(111) has fewer structural defects than it does on Cu(100).

I. INTRODUCTION

Graphene growth on copper foils is attractive as a low-cost and simple method to synthesize high-quality graphene.[1] Understanding how substrate morphology and crystallinity affect defect formation is crucial to improving film quality. This effort is complicated in the case of Cu foils, because their crystallographic texture varies with the manufacturing process — cold-rolled foils recrystallized by annealing have marked (100) texture.[2] But other processes lead to low-energy (111) surfaces in foils[3-4] and films. The detailed morphology (*e.g.*, the distribution of surface steps) of foil surfaces also depends on preparation details. We previously reported that graphene grown by depositing C on (100) grains of Cu foils has substantial in-plane rotational disorder.[2] The in-plane orientations are around two crystallographically equivalent Cu directions, a consequence of placing the sixfold graphene on the fourfold (100) substrate. The range of orientations around the two Cu directions is large, ~ ± 7.5°, as illustrated in Fig. 1(a). Furthermore, each nucleation site typically generates four graphene crystals, each with a different in-plane orientation. A large density of rotational boundaries results when these misoriented islands grow and impinge, lowering film quality.

In contrast, relatively little is known about the origin of such mosaicity for graphene grown on Cu(111) surfaces. A postgrowth analysis by scanning tunneling microscopy (STM)[5] has shown that a relatively high density of rotational domain boundaries exists in graphene grown by ethylene decomposition. Whether these rotational domains are generated in the initial nucleation events[2, 6] or during subsequent growth[7] is not clear.

Insight into the atomic growth mechanisms can be obtained by analyzing the shapes of growing islands. For example, the distinctive four-lobed islands grown on Cu(100) in ultrahigh vacuum (UHV) arise from a combination of simultaneously nucleating several rotational domains and a mechanism of carbon atom attachment that depends on the orientation of the graphene edges.[2] Although such lobed islands on Cu foils are also observed in chemical vapor deposition (CVD),[1] several groups[8-10] have shown that hexagonal, single-crystal islands can also form on Cu foils. With low-energy electron microscopy (LEEM) and selected-area low-energy electron diffraction (LEED), Li *et al.* observed sixfold, snowflakelike islands on a (100)-oriented grain in a foil.[11] These findings suggest that graphene may grow on Cu by several mechanisms, depending on the synthesis conditions.

Here, we gain insight into the growth mechanism by using LEEM to observe graphene growing on a Cu(111) single crystal exposed to a flux of elemental carbon. We find that graphene first nucleates inhomogeneously at defects such as substrate steps,[12] step bunches, and impurities on practical Cu(111) surfaces. Graphene's in-plane alignment and island morphology strongly depend on substrate temperature. At low temperatures (< 700 °C), the islands are highly dendritic. Within each island, the in-plane orientation changes substantially over submicron length scales. At high growth temperatures (> 900 °C), the islands are more compact but still have distinct lobes. The graphene lattices of all lobes are closely aligned with the Cu(111) lattice, unlike islands on Cu(100).[2] We show that these dendritic shapes occur because the growth rate is limited by surface diffusion of a C species. In stark contrast, the growth rates on Ir(111) and Ru(0001) substrates are limited by large energetic barriers of attaching the growth species.[13] We observe that new rotational boundaries can be generated as graphene sheets grow across Cu step bunches. Consistent with this mechanism, islands nucleated at large bunches of Cu steps tend to be polycrystalline, while those nucleated on flatter regions are single crystals. Thus, crystallographic alignment is also strongly affected by the Cu surface morphology. Overall growth on smooth Cu(111) surfaces can produce graphene films closely aligned to a single in-plane orientation, in contrast to the large rotational disorder found for Cu(100) substrates under the same growth conditions.[2]

## II. EXPERIMENT

The Cu(111) single crystal was cleaned initially by annealing in a tube furnace at 900 °C for 12 hours in an Ar-$H_2$ mixture at atmospheric pressure. Before each growth in the LEEM, the crystal was exposed to $1\times10^{-7}$ torr oxygen at 950 °C to remove carbon and then sputtered in $O_2$. Following annealing at 500 °C, several more cycles of Ar sputtering and annealing were performed. The substrate temperature was measured by a thermocouple spot welded to a molybdenum washer pressed against the backside of the crystal. Since hydrocarbons like ethylene do not decompose easily on Cu under UHV conditions, we deposited carbon from a graphite rod heated by an electron beam. LEEM images were acquired during graphene growth. After cooling to room temperature, the orientation of individual graphene islands was determined by selected-area LEED and dark-field LEEM images obtained from first-order graphene diffraction spots.

## III. RESULTS AND DISCUSSION
### A. Island nucleation

Figure 2(a) shows a LEEM image obtained shortly after graphene islands nucleated on Cu(111) at 815 °C. Most islands nucleated at atomic steps or step bunches on the Cu surface, similar to other transition metals.[14] In addition, some islands nucleated at impurity clusters. Thus, the initial nucleation on the stepped and terraced Cu(111) surface is clearly heterogeneous in nature. To further probe the processes that govern nucleation, we increased the C flux after the initial nucleation events, inducing secondary nucleation. Figure 2(b) shows that the new islands formed mainly at positions equidistant from the original islands. This signature of diffusion-limited growth (see Sec. III. E) is observed only when heterogeneous nucleation does not dominate. Thus, the distribution of substrate defects is not the only factor that determines the distribution of nucleation sites.

In past work, we used changes in electron reflectivity to measure quantitatively the concentration of C adatoms on Ru(0001) and Ir(111)[13] surfaces during graphene growth. Similar measurements on Cu(111) detected no measureable changes in electron reflectivity from the start of C deposition until island nucleation, when the maximum concentration is expected. We estimate the surface carbon concentration on Cu(111) to be $<1\times10^{-3}$ monolayer during growth. The low-energy binding site of single C atoms is calculated to be underneath the first Cu layer.[15] If deposited C can easily reach this low-energy site, little C will exist as surface adatoms.

B. Dependence of crystallinity on nucleation process

The in-plane orientation of the islands depicted in Fig. 2 was determined using selected-area diffraction. About half of the islands gave a single set of sixfold graphene diffraction spots closely aligned to the Cu spots (see the representative pattern in Fig. 3(b)). These islands, colored red (online) in Fig. 3(a), are single crystals whose lattices align with the Cu lattice (see Fig. 1(b)). The other half of the islands is polycrystalline, colored blue (online) in Fig. 3(a). Their diffraction patterns contain more than one set of graphene spots rotated with respect to one another. Within a 2-μm-diameter region of a single island (Fig. 3(c)), we found up to seven different in-plane orientations (grains).

A strong correlation exists between the crystallinity of the islands and the time at which they nucleated. From the real-time imaging (see supplementary material movie 1[16]) we know that 10 of the 11 single-crystal islands in Fig. 3(a) were secondary nuclei, forming after increasing the C flux. This observation does not imply that high flux helps form single-crystal islands. Rather, the nature of the nucleation site affects whether the island is a single crystal. The nucleation sites in defective, rough regions of the surface preferentially form polycrystals. Single crystals preferentially form at the less defective regions when higher C concentrations cause secondary nucleation. This suggests that suppressing the first type of nucleation will yield graphene films with fewer rotational boundaries. Gao *et al.*[5] also suggested that decreasing defect densities on Cu(111) leads to more uniform graphene growth. Thus, higher-purity and low-step-density Cu(111) surfaces are desirable, reinforcing the importance of the pretreating the substrate to minimize native oxide and morphological defects.[17] Next, we show that higher growth temperature also helps eliminate large-angle rotational boundaries within films.

C. Dependence of crystallinity on growth temperature

Figure 4 shows that graphene islands grown at lower temperature, 690 °C, are markedly more dendritic than those grown at 815 °C (see Fig. 3). The islands are preferentially elongated along bunches of substrate steps, showing that the bunches have a stronger influence at lower temperatures (see supplementary material movie 2[16]).

Typical morphologies of graphene grown at 900, 950 and 975 °C are shown in Figs. 5(a), 6(a) and 6(b), respectively. Compared with lower-temperature growth (690 °C in Fig. 4 or 815 °C in Fig. 3), these islands are more compact and have smoother edges. At the highest growth temperatures (Fig. 6(b)), the island edges become more distinctly faceted.

Temperature also has a strong effect on crystallinity. Diffraction patterns from 0.5-μm-diameter areas of individual islands grown at 690 °C, (see Fig. 4(b)), have multiple sets of graphene spots. Thus, the islands are composed of small rotational domains. Most islands (> 90%) grown at 900 °C give a single set of diffraction spots that are sharp or have arcs that span < 3°. The sharp spots or arcs are either closely aligned with (see Fig. 5(b)) or rotated by a few degrees (see Fig. 5(c)) from the Cu spots.

More information about the spatial distribution of graphene's in-plane orientation is offered by dark-field LEEM. Five dark-field images were obtained at angular separations of 1.5° along the arc of first-order diffraction spots of graphene grown at 900 °C. Each image was assigned a different color, whose saturation is proportional to the image intensity. The composite of the five images, shown in Fig. 7, provides a real-space map of in-plane orientation. Over the 20-μm field of view, all the graphene islands are aligned within ±3° of the Cu(111) lattice. Thus, high-temperature growth has the clear benefit of aligning most graphene islands to a single in-plane orientation. Figure 7 shows that there is still some rotational disorder, roughly ±1.5°, within individual islands. This disorder exists even though real-time observation (see supplementary material movie 3[16]) showed that each island grew from a single nucleation site. (The dark-field aperture had an ~ 2° acceptance angle, leading to some overlap among the individual images and making precise determination of the boundary sharpness challenging.) We next discuss how rotational disorder develops during growth of graphene sheets.

D. Origin of the mosaicity

A basic question is whether the rotational domains within single islands such as those in Figs. 3 and 4 are generated during the initial nucleation event or during subsequent growth. The former occurs on Cu(100), where multiple rotational domains arise from nucleation at a defect.[2, 18] The latter occurs on Ir(111), where rotational domains are observed to form on the edges of expanding graphene sheets.[7] At higher temperatures, the rotational domains on Cu(111) are large enough to allow their formation to be monitored, as shown in Fig. 8 (and see supplementary material movie 4[16]). As with Ir, new rotational domains can occur as an island expands. The image sequence in Fig. 8(a)-(c) shows a graphene sheet advancing toward a Cu step bunch, marked by the dotted red line. The growth velocity decreased at the step bunch. After graphene crossed the bunch, selected-area diffraction (Fig. 8(d)) showed that the graphene below the Cu step bunch (yellow circle in Fig. 8(c)) was rotated 21° from the graphene above the bunch (green circle in Fig. 8(c)). Thus, the step bunch led to a rotational boundary in the island, as sketched in Fig. 8(e). Zhao *et al.* also observed a change of graphene orientation across Cu steps using STM.[10]

We believe that this mechanism is a general source of rotational disorder in graphene growth on Cu(111). Since the initial island nucleation occurs in rough regions of the surface, this effect could explain why the first-nucleated islands in Fig. 2 tend to be polycrystalline. We also suggest that the substrate steps introduce rotational disorder into graphene islands during high-temperature growth (Fig. 7). The tendency to introduce rotational boundaries during growth differentiates Cu(111) from other surfaces. For example, graphene sheets can grow without changing orientation across boundaries between rotationally misoriented Ru(0001) grains[19] and even across different facets of Ni grains.[20]

E. Diffusion-limited growth: experiment and modeling

In this section, we show that the growth rate on Cu(111) in our experiments is limited by surface diffusion. The dendritic shapes in Fig. 4(a) are suggestive of the instabilities that occur during diffusion-limited growth.[21] Islands grown at higher temperatures are more compact but still have distinct lobes (see Figs. 5 and 6). At first glance, the lobed islands resemble growth on Cu(100). In that system, however, the asymmetric growth shapes of the four-lobed islands were interpreted in terms of orientation-dependent attachment barriers. That is, the rate-limiting barrier of attaching C adatoms varies with the in-plane orientation of the graphene edge. But such attachment asymmetries cannot cause the shapes we find on Cu(111) — an attachment barrier for growth on Cu(111) would be sixfold symmetric and the kinetic growth shape[22] would be a compact island, consisting of six edges in the slow-growth directions. In contrast, some islands on Cu(111) are distinctly noncompact, having six lobes. The significant deviations from perfect sixfold symmetry likely result from the tendency of the lobes to follow the directions of the underlying Cu step bunches. Indeed, Fig. 9(a) shows a reasonably symmetric six-lobed island that is characteristic of growth in regions with low step densities. (The island nucleated at an isolated screw dislocation and is a single crystal, as the diffraction pattern in Fig. 9(b) shows.) We next look more closely at the nucleation process and examine how individual islands grow. We find that diffusion-limited growth occurs even at the higher temperatures.

In diffusion-limited growth, the growth species has significant concentration gradients across the surface. Local concentration maxima occur in regions that are (roughly) equidistant from neighboring islands. These maxima should be positions of enhanced nucleation. For example, they should determine the points of secondary nucleation in experiments like Fig. 2(b), where the flux was increased after the initial nucleation events. To test this hypothesis, we numerically solved the two-dimensional diffusion equation for the experimental configuration of growing islands shown in Fig. 10(a). The carbon concentration $c$ on the terraces between islands was determined assuming a constant incident flux. Each island boundary was assumed to be a perfect sink (*i.e.*, $c = 0$ there). The edge of the field of view was taken as a perfectly reflecting boundary. Figure 10(b) shows the result, where the grayscale intensity is proportional to concentration $c$. Abruptly increasing the experimental flux gave the secondary nucleation shown in Fig. 10(c). The red crosses on Fig. 10(b) mark the positions of the new experimental nuclei on the calculated concentration profile. There is a clear tendency for enhanced nucleation near the predicted concentration maxima. (The average $c$ at the positions of new nucleation is ~30% higher than the average $c$ within the field of view.) However, clear exceptions occur. Presumably, these arise from the already discussed fact that nucleation is not homogeneous and is to some extent determined by the position of surface defects. Nevertheless, the preferential nucleation in regions with a high predicted $c$ is suggestive that significant concentration gradients exist on the Cu(111) surface.

We next confirm directly that concentration gradients influence observed island shapes, i.e., that morphological instabilities exist. To do so, we measured the flux to the edge of several islands during growth and compared to predictions from calculated diffusion gradients. Figure 11(a) shows the starting experimental island configuration. Figure 11(b) gives the configuration 61 s later. In Fig. 11(c), the difference of Figs. 11(b) and 11(a), the width of the line surrounding each island measures the flux to each segment of island edge. Figure 11(d) shows the flux calculated from the diffusion equation, set up as in Fig. 10(b). There is a striking similarity with the experimentally determined configuration. For example, in both theory and experiment, the corners of the star-shaped island marked "A" in Fig. 11(c) grow roughly five times faster than the depressions. (This difference is what causes flat interfaces to become unstable.) This observation provides strong evidence that growth is indeed diffusion limited.

However, factors other than diffusion gradients also must affect island shapes. In simple diffusion-limited aggregation, the shapes would always be fractal. But here, the shapes change with temperature (see Figs. 4–6). Furthermore, diffusion should be isotropic on a (111) surface. But here, the islands at higher temperature are clearly sixfold symmetric. An explanation of the temperature dependence is that there must be thermally activated processes that smooth rough step edges, such as edge diffusion or detachment/reattachment at step edges. A possible explanation of the sixfold asymmetry builds on this observation: small islands formed immediately after nucleation are hexagonal because of fast edge diffusion, for example.[23] As these shapes expand, the six corners grow more quickly, initiating the observed sixfold shape asymmetry. Figure 9(c) gives a schematic illustration.

Interestingly, compact hexagonal shapes have been observed in high-pressure CVD by several groups.[8-10] This observation indicates that growth under these conditions is not surface-diffusion limited. As pointed out by Bhaviripudi *et al.*[24], CVD may be limited by gas-phase diffusion. Also, the carrier and carbon-source gases in CVD suppress Cu evaporation so that a higher temperature can be employed compared to that used with UHV growth. At high temperatures in our UHV experiments, the surface morphology is evolving quickly due to sublimation, causing large step bunches to collect at graphene edges, as seen in Fig. 6(b). Rearranging these step bunches is likely difficult, impeding the processes that lead to hexagonal shapes.

Diffusion-limited growth on Cu(111) is surprising, because growth on substrates like Ru(0001) is limited by the energetic barrier of attaching the growth species, not the rate of surface diffusion.[13] The existence of such attachment barriers is easy to understand, because single C atoms must break bonds with the metal substrate before binding with the graphene. Surprisingly, growth at similar temperatures on Cu(100) is attachment limited,[2] even though carbon adatoms on this surface should diffuse more slowly than on the close-packed (111) face.[25] One explanation for diffusion-limited growth on Cu(111) is that, unlike most other metals, the low-energy binding site of single C atoms is underneath the first Cu layer.[15] Movement of this atom would presumably require its thermal excitation into an adatom, leading to slow surface diffusion. However, this process might also be expected to give a significant attachment barrier; so the detailed explanation is still lacking.

IV. CONCLUSIONS

Graphene islands grown on Cu(111) are often polycrystalline. The degree of polycrystallinity depends strongly on growth temperature, surface roughness, and surface defects. Cu(111) step bunches lead to rotational disorder in two ways. First, they cause islands to be nucleated with different in-plane orientations (see Figs. 2 and 3). Second, step bunches can generate new rotational boundaries as islands expand (see Fig. 8). Thus, fewer step bunches lead to fewer rotational boundaries, consistent with the work of Zhao *et al.*[10] High-angle rotational boundaries can be minimized using higher growth temperatures, yielding larger graphene grains that contain only low-angle (< ±3°) rotational boundaries (see Fig. 7). In contrast, even under ideal conditions, two orientations of graphene nucleate for symmetry reasons on Cu(100) (see Fig. 1(a)). Thus, the precise alignment to a single Cu(111) direction achieved under optimized conditions is a significant advantage over Cu(100). The growth rate is limited by diffusion of a C species along the Cu(111) surface, unlike Cu(100) and other metals. The observation of island shapes becoming more compact at higher growth temperatures shows that an additional diffusion process, such as edge diffusion, smoothes the island edges. This equilibration process may also lead to higher-quality graphene by healing any point defects created during growth.

ACKNOWLEDGMENTS

Work at Sandia and the Lawrence Berkeley National Laboratory was supported by the Office of Basic Energy Sciences, Division of Materials Sciences and Engineering, U. S. Department of Energy, under Contracts No. DE-AC04-94AL85000 and No. DE-AC02-05CH11231, respectively. J.M.W. acknowledges support from the National Science Foundation Graduate Research Fellowship Program.

‡ Author to whom correspondence should be addressed. mccarty@sandia.gov

Figure Caption

**FIG. 1.** (Color online) (a) Alignment of graphene on Cu(100). Graphene grows with a wide spread of in-plane orientations around two symmetry-equivalent Cu(100) directions. (b) Alignment of graphene on Cu(111). Under optimized conditions, graphene grows closely aligned to a single in-plane orientation.

**FIG. 2.** LEEM images of graphene islands growing at 815°C on Cu(111) (a) after initial nucleation and (b) after secondary nucleation following an increase in the carbon flux. Graphene is bright while dark stripes are Cu step bunches. The field of view is 20 μm.

**FIG. 3.** (Color online) (a) LEEM image of the region in Fig. 2 after continued growth at 815 °C. Field of view is 14.5 μm. Red islands are single crystals rotationally aligned within 4° of the Cu lattice and mainly nucleated after increasing the carbon flux. Blue islands are polycrystalline. (b-c) Typical LEED patterns of the red and blue islands, respectively, at 50 eV.

**FIG. 4.** (a) Dendritic graphene grown at 690 °C, suggesting a diffusion-limited process. The field of view is 7 μm. (b) LEED pattern (44 eV) from a 0.5-μm-diameter region of an island, showing that it is polycrystalline.

**FIG. 5.** (Color online) (a) LEEM image of graphene grown at 900 °C. The field of view is 46 μm. (b-c) LEED patterns (50 eV) taken inside the areas enclosed by the red and green lines, respectively. The electron beam size is ~ 2 μm. Most islands are aligned with the Cu lattice within a small rotation.

**FIG. 6.** LEEM images of faceted islands grown at (a) 950 °C (20-μm field of view) and (b) 975 °C (14.5-μm field of view).

**FIG. 7.** (Color online) Dark-field analysis of graphene grown at 900 °C. The image is a composite of five dark-field micrographs obtained in 1.5° rotational increments from the $Cu[11\bar{2}]$ direction at 0°. The saturation of each color reflects the degree graphene is aligned to each angle. The field of view is 20 μm.

**FIG. 8.** (Color online) (a-c) Sequence of LEEM images showing a graphene island growing at 893˚C (9 μm × 4 μm). The red dotted line marks a Cu step bunch. (d) LEED patterns from the areas within the yellow and green circles in (c). The graphene below the red dotted line is rotated with respect to the graphene above the line. (e) Schematic depicting change of island orientation (green to yellow) that arises when an island grows across a step bunch.

**FIG. 9.** (Color online) (a) Graphene island with six lobes on a Cu screw dislocation (14.5-μm field of view). (b) LEED pattern (50 eV) showing that the island is a single crystal. (c) Schematic illustration of a compact (dark) island evolving into a six-lobed shape (gray) during diffusion-limited growth.

**FIG. 10.** (Color online) (a) LEEM image showing an array of graphene islands during growth at 894 ˚C (46-μm field of view). (b) Carbon concentration calculated by the model described in the text. The concentration is low (dark) near the islands and the highest (bright) at regions farthest from any island. (c) Experimental configuration after secondary islands nucleated following an increase in the C flux. The red crosses in (b) mark the positions of the new islands in (c).

**FIG. 11.** (a-b) LEEM images separated by 61 s during growth at 893 ˚C (20-μm field of view). (c) Difference between (b) and (a), where the bright strips show the incremental growth. (d) Flux to the graphene edges computed by solving the diffusion equation for a uniform deposition flux. The grayscale intensity is proportional to the flux to the island edges.

Fig. 1

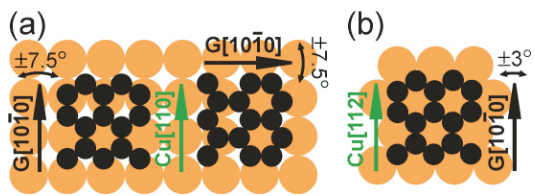

Fig. 2

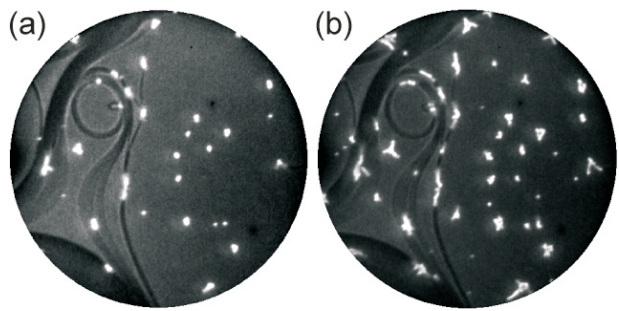

Fig. 3

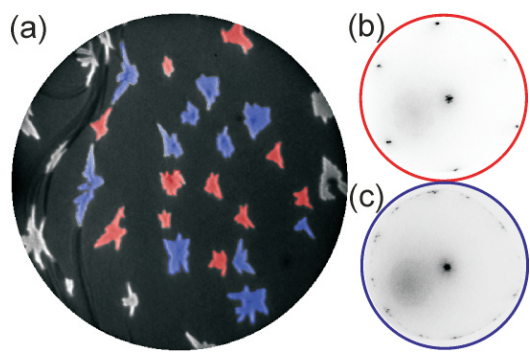

Fig. 4

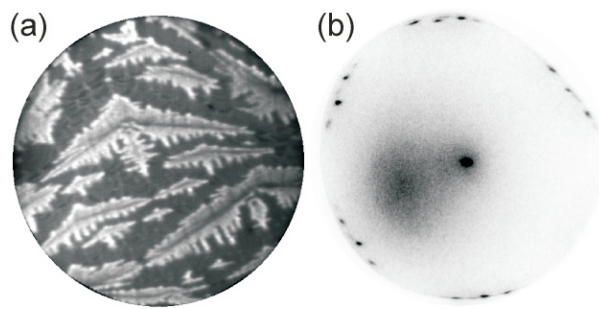

Fig. 5

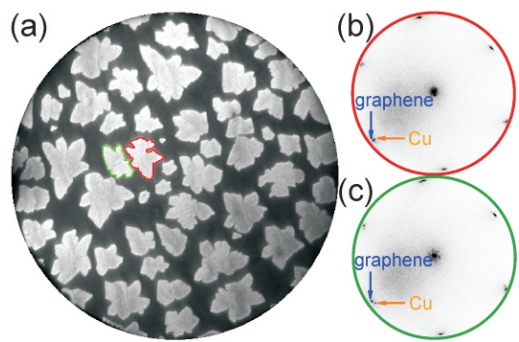

Fig. 6

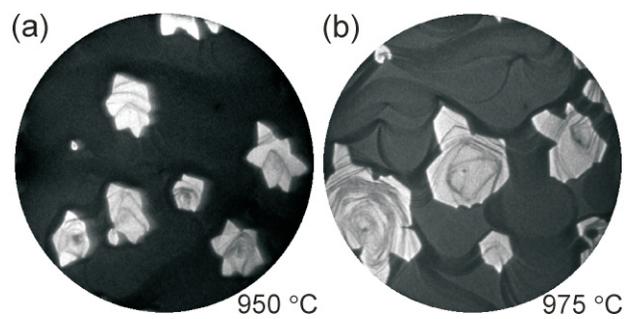

Fig. 7

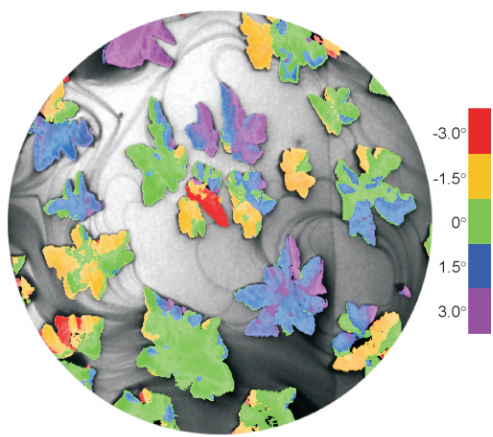

Fig. 8

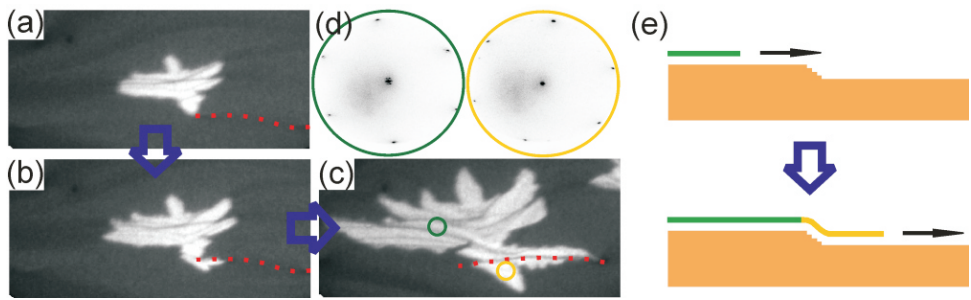

Fig. 9

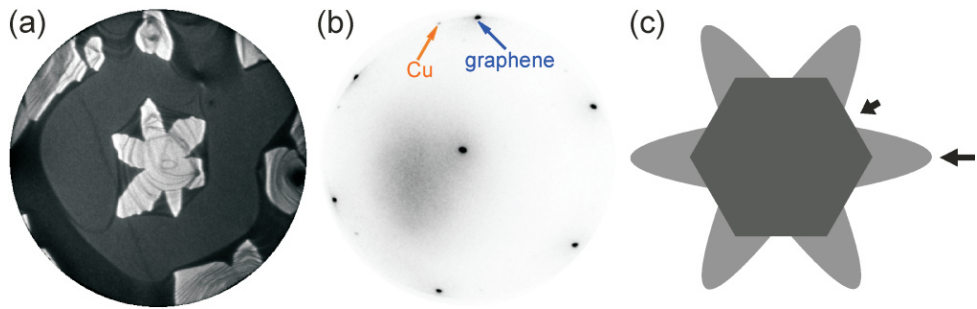

Fig. 10

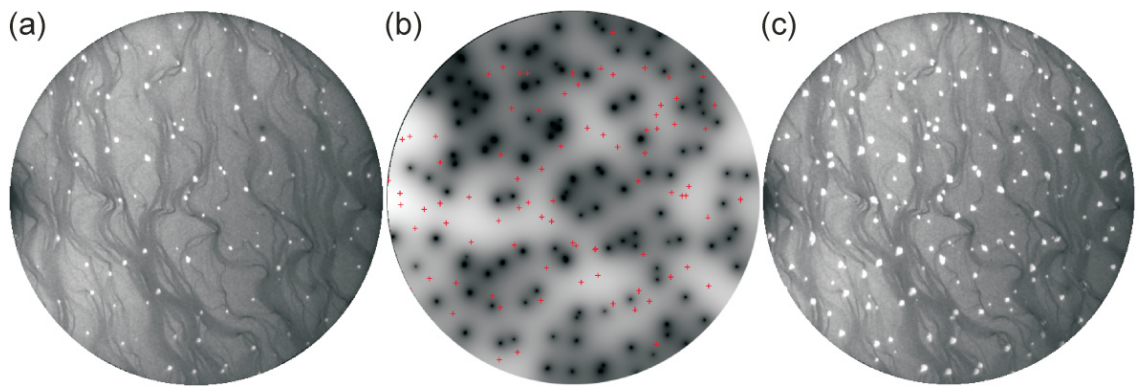

Fig. 11

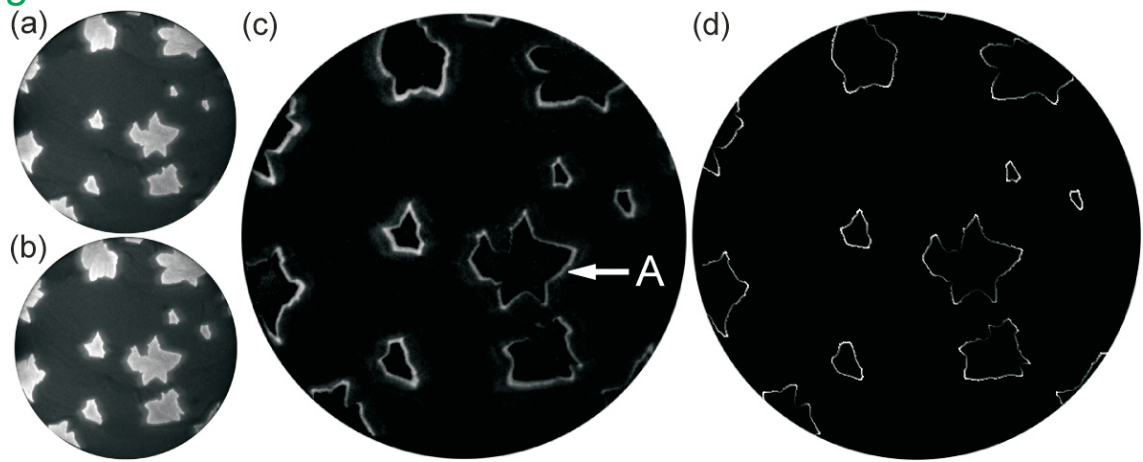